\documentclass[aps,prl,twocolumn,showpacs,superscriptaddress,groupedaddress]{revtex4-1}
\usepackage{graphicx}
\usepackage{subfigure}
\usepackage{dcolumn}
\usepackage{bm}
\usepackage{amssymb}
\usepackage{multirow}

\newcommand\ab{{\langle{\bf A} \cdot{\bf B}}\rangle}
\newcommand\bfv{{\bf v}}
\newcommand\bfb{{\bf B}}

\def\bfv{{\bf v}}

\def\bfb{{\bf b}}

\def\emfb{\overline{\mbox{\boldmath ${\cal E}$}} {}}
\def\beq{\begin{equation}}
\def\ee{\end{equation}}

\begin{document}

\title{A study of large scale dynamo growth rates from numerical simulations  and implications for mean field theories}
\author{Kiwan Park}
 \altaffiliation{Department of Physics and Astronomy, University of Rochester, Rochester, New York US, 14627}
\author{Eric G. Blackman}
\affiliation{Department of Physics and Astronomy, University of Rochester, Rochester, New York US, 14627}
\author{Kandaswamy Subramanian}
\affiliation{IUCAA, Post bag 4, Ganeshkhind, Pune 411 007, India}
\date{\today}

\begin{abstract}
Understanding large scale  magnetic field growth in   turbulent plasmas in the  magnetohydrodynamic limit  is  a goal  of magnetic dynamo theory.
In particular,  assessing how well large scale helical  field growth and saturation in simulations matches that predicted by existing theories is important for progress.  Using numerical simulations of isotropically forced  turbulence without large scale shear with the implications, we focus on several aspects of this comparison that have not been previously tested:
(1) Leading mean field dynamo theories which break the field into large and small scales
 predict that large scale helical field growth rates are determined by the difference between kinetic helicity and current helicity with no  dependence on the non-helical energy in small scale magnetic fields.
  Our simulations show that the growth rate of the large scale field from fully helical forcing is indeed unaffected by the presence or absence of  small scale magnetic fields amplified in a  precursor non-helical dynamo.  However,  because the precursor non helical dynamo in our simulations produced  fields that were strongly sub-equipartition with respect to the kinetic energy,   we cannot yet rule out the potential influence of  stronger non- helical small scale fields.
 (2)  We have identified two features  in our simulations which  cannot be explained by the most minimalist versions of two-scale mean field  theory:   (i)   fully helical small scale forcing produces significant  non-helical large scale magnetic energy and (ii)  the saturation of the large scale field growth  is time-delayed with respect to what minimalist theory predicts. We comment on desirable generalizations to the theory in this context and future desired work.
\end{abstract}

\maketitle

\begin{figure}

\caption{(a) For $\eta=\nu=0.006$. The lines in large plot show $H_1$ growth in linear scale  the small inset box shows $H_1$ in Log scale.  The solid lines show the growth in the  ``$HF$  after $NHF$'' phase  but with the  $NHF$ removed, thus starting the x-axis at origin (${t'}_{006}=t-714$, see text).  The dashed lines show the case of  $HF$ without $NHF$ plotted twice in the large plot. The left dashed line is the original in simulation time,  and the right is time-shifted by($t''_{006}=t'_{006}+275$) for  comparison with the solid line.  Note that the time shifted dashed curve matches the solid curve well. (b) $\eta=\nu=0.001$. Same as (a) but with the analogous time shifts $t'_{001}=t-584.7$ and $t''_{001}=t_{001}'-128$ (see text).  (c) All 4 simulation curves  shown on the same plot  in two sets: from left to right, the first 4 curves are the same as those plotted in parts (a) and (b) as the legend indicates. These 4 curves are then all time shifted to overlap
near
$t=1000$
on the time axis in order to aid the visual comparison of all simulations for the two different $\eta=\nu$ cases.}

\subfigure[]
   {
     \includegraphics[width=8.2cm]{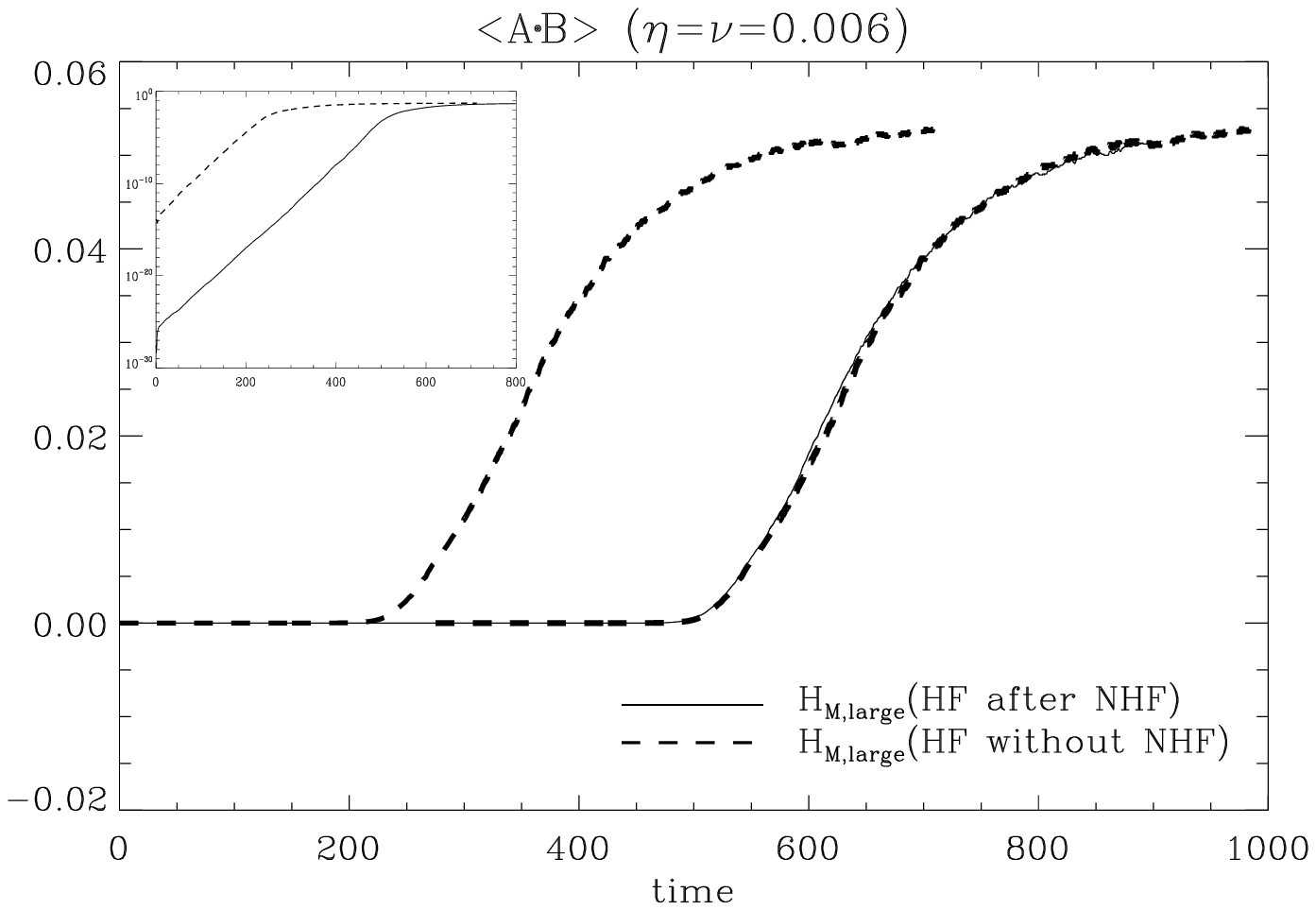}
     \label{hl006}
               }\,
   \subfigure[]
   {
     \includegraphics[width=7.6cm]{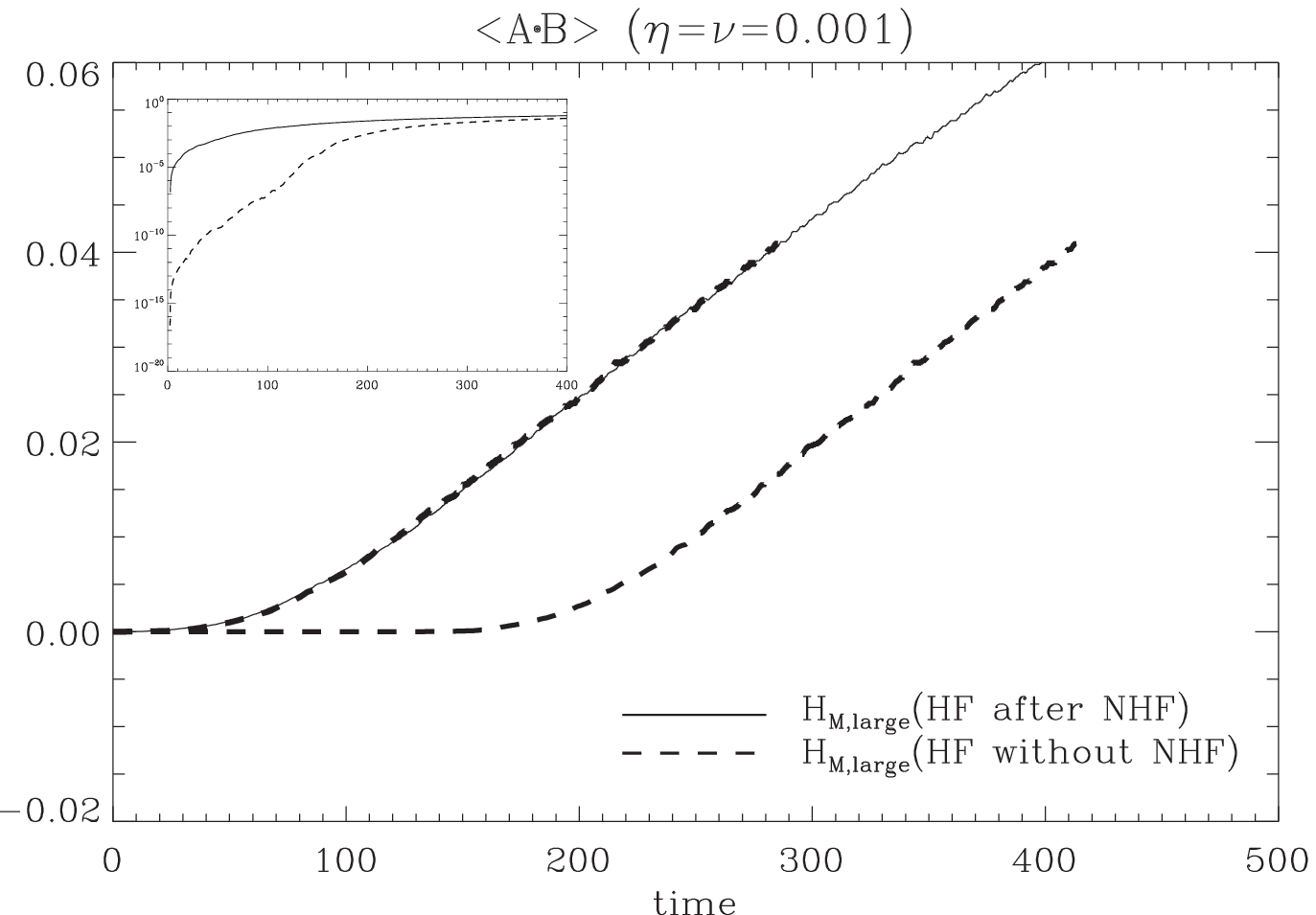}
     \label{hl001}
               }\,
   \subfigure[]
   {
     \includegraphics[width=8.4cm]{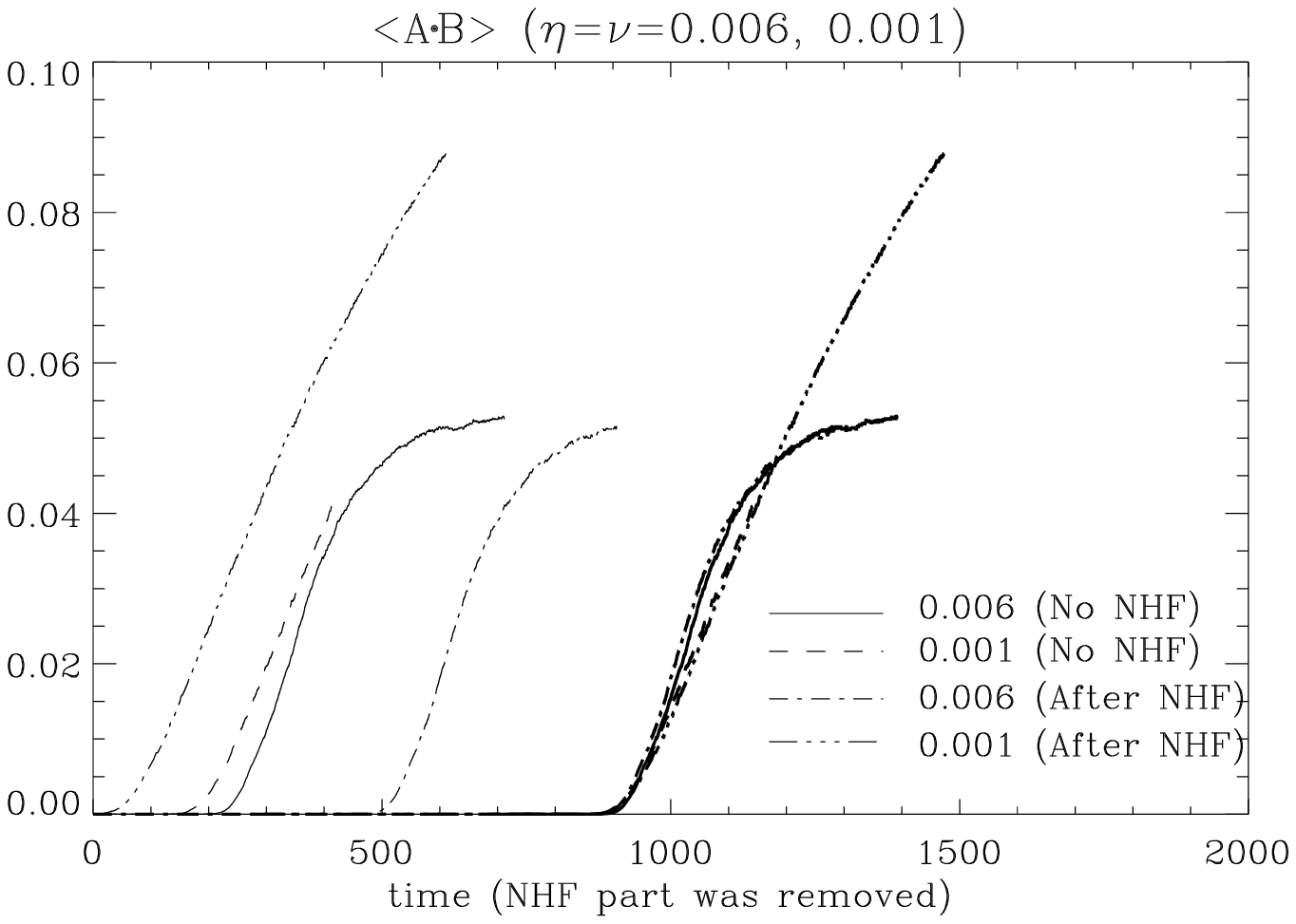}
     \label{hl006001}
               }\,

\end{figure}

\section{Introduction}
The origin of magnetic fields in  planets, stars, galaxies, accretion disks   (\cite{1996ARA&A..34..155B}, \cite{2010AN....331..101B}, \cite{2009SSRv..144...87B}, \cite{2005PhR...417....1B}, \cite{1978mfge.book.....M})
comprises an ongoing enterprise of investigation. Understanding  mechanisms by which large scale  fields can be amplified  from weak seed fields by the underlying velocity flows has been a particular long standing  goal.
The specific manifestation of magnetic amplification in these  systems comprises a diverse set of boundary conditions, viscosities, thermal conduction coefficients, and magnetic diffusivities as described in these references. But conditions are such that the Magnetic Reynolds numbers $Re_M$(=$UL/\eta$, `U', `L': typical velocity and length scale of the flow, `$\eta$': magnetic diffusivity) are  typically high enough for  the magnetohydrodynamic(MHD) approximation to be useful.   Moreover, while the ratios of hydrodynamic to magnetic diffusivities vary widely from being much greater than unity in galaxy plasmas to being much less than unity in planetary cores,  the hydrodynamic Reynolds numbers are  high enough that the flow are turbulent.

When it is possible to identify a dominant energy containing scale of turbulence (typically the forcing scale),   a natural distinction arises between two classes of dynamos:  Small scale dynamos (SSDs) describe the  amplification of fields at or below the dominant turbulent eddy scale (\cite{1968JETP...26.1031K}, \cite{2004ApJ...612..276S})
  whereas
large scale dynamos (LSDs) describes the amplification of fields on spatial or temporal scales  larger than the those of the underlying forcing.
The ingredient that distinguishes standard velocity driven LSDs from SSDs is the presence of a natural pseudo scalar in the turbulence, such as  kinetic helicity. (\cite{1976JFM....77..321P}, \cite{1981PhRvL..47.1060M}).
Proof of principle of large scale field growth has also been recently emerging in the laboratory e.g.
\cite{2012PhRvL.108n4501G}. But experiments are presently limited to the kinematic regime, where the field amplification is so weak that it does not much influence the velocity flows.
 An important aspect of astrophysical dynamos  that is accessible in numerical experiments  but not in laboratory experiments is the nonlinear evolution when the magnetic field grows large enough to modify the driving velocity flows.

Objects such as galaxy clusters  have no natural sources of  helicity and would be dominated by SSDs, but most astrophysical rotators such as stars and galaxies  typically sihave  the ingredients needed to drive contemporaneous LSD and SSD (see e.g. \cite{2005PhR...417....1B}, \cite{2010AN....331..101B}). A long-standing  concern in dynamo theory \cite{1981coel.book.....P}, \cite{1992ApJ...396..606K} has been the extent to which the  presence of  rapidly growing SSDs influences  LSDs.
A detailed comparison of the exact growth rates for an LSD with and
without a precursor SSD has not been carried out and is the goal of this paper.

Most of the work on LSDs and SSDs has been separate,  but minimalist simulations of
the two types of dynamos can be studied in similar set-ups, differing only  in the nature of the forcing.
Starting with a triply periodic box and a weak seed magnetic field,   the $\alpha^2$ dynamo in a periodic box was simulated \cite{1981PhRvL..47.1060M}, \cite{2001ApJ...550..824B} by forcing with kinetic helicity at wavenumber $k_f=5 k_{min}$  The large-scale ($k=1$) field grows and saturates as  predicted from spectral  Ref.~\cite{1976JFM....77..321P} and two-scale theories \cite{2002PhRvL..89z5007B}:  Driving with kinetic helicity initially  grows a large scale helical magnetic field but the near
conservation of magnetic helicity leads to a compensating small scale
magnetic (and current) helicity of opposite sign to that on the large scale. This counteracts the
kinetic helicity driving and quenches the LSD.  In these theories, the non helical magnetic energy in the smaller scale does not play a role in the growth or saturation.

Complementarily, there have  been many  SSD simulations in which the same triply periodic box is used but the  forcing is non-helical, (e.g. \cite{1981PhRvL..47.1060M}, \cite{2003ApJ...597L.141H},  \cite{2004PhRvE..70c6408H},  \cite{2004ApJ...603..569M},  \cite{ 2005ApJ...626..853M},  \cite{2013MNRAS.429.2469B},  \cite{2004ApJ...612..276S}).
 SSD simulations without large scale shear show  that the total magnetic  energy is amplified exponentially in the kinematic regime,  as analytically predicted  \cite{1968JETP...26.1031K}, \cite{1992ApJ...396..606K} and evolves to saturate  at  10-30\% of equipartition   with the total kinetic energy  in fully nonlinear regime.  For large magnetic Prandtl number $Pr_M=\nu/\eta$, where $\nu$ and $\eta$ are the kinematic
viscosity and magnetic diffusivity, the spectrum shows super-equipartition values on small scales
even when the total magnetic energy is sub-equipartition
 \cite{2003ApJ...597L.141H}, \cite{2004PhRvE..70c6408H},  \cite{ 2002ApJ...567..828S}, \cite{2004ApJ...612..276S}.
However, synthetic Faraday rotation observations of a SSD system  show  that strong small scale  structures would not  dominate the  observed structures (\cite{2013MNRAS.429.2469B}),
nor is there observational evidence for them.


The goals here are  to (1)  assess whether a precursor  SSD affects the subsequent growth of an LSD
and (2)  to evaluate how well simple two-scale theories fit the results of the resultant growth rates
and evolution of the large scale field. In second section, we discuss the simulation set up and the methods used. In third section we discuss the results of the simulations. We interpret these results and their implications for theoretical modeling in fourth section and conclude.

\section{ Problem to be solved and methods}

We carry out two types of  simulations  for two different $R_M$ all for Prandtl number unity. For simulation type 1, we start the simulation with fully helical forcing HF in the momentum equation which leads to    $\alpha^2$ dynamo growth of the large scale magnetic field. For simulation type 2, we start with non-helical forcing (NHF) in the momentum equation and then only later  turn on the fully helical forcing. For the high $\eta$ case, the NHF phase does not amplify the small scale field because it is subcritical to the SSD. For the low $\eta$ case, the system is  unstable to SSD growth and the total field grows exponentially.  Our runs allow us to compare the LSD of the HF phase for three different cases:   NHF $\rightarrow$ HF but without SSD;  NHF $\rightarrow$ HF with  SSD and
HF without any NHF phase.

We use the high-order finite difference Pencil Code(\cite{2001ApJ...550..824B}) and  the message passing interface  (MPI) for parallelization.
The  equations solved   are the compressible MHD equations given by
 \begin{eqnarray}
\frac{D \rho}{Dt}&=&-\rho \nabla \cdot {\bf v}\nonumber\\
\frac{D {\bf v}}{Dt}&=&-c_s^2\nabla \mathrm{ln}\, \rho + \frac{{\bf J}\times {\bf B}}{\rho}+\frac{\tilde \mu}{\rho}\big(\nabla^2 {\bf v}+\frac{1}{3}\nabla \nabla \cdot {\bf v}\big)+{\bf f}\nonumber\\
\frac{\partial {\bf A}}{\partial t}&=&{\bf v}\times {\bf B} -\eta\,{\bf J}
\label{MHD equations in the code}
\end{eqnarray}
where  $\rho$ is the density; $\bf v$ is the velocity; $\bf B$ is the magnetic field; $\bf A$ is the vector potential; ${\bf J}$ is the current density;  $D/Dt(=\partial / \partial t + {\bf v} \cdot \nabla$) is the advective derivative; $\eta$ is the magnetic diffusivity; $c_s$: sound speed; and $\bf f$ is the forcing function.
The ratio $\nu = {{\tilde \mu}\over \rho}$ is the mean kinematic viscosity for given dynamic viscosity $ \tilde \mu$.
Since $c_s$ is constant in the simulations,  dimensionless units are constructed such that velocities are expressed in units of $c_s$ and magnetic fields expressed in units of $c_s( \mu_0 \rho_0)^{1/2}$, where
$\mu_0$ is the magnetic permeability and $\rho_0$ is the initial density. These constants are then set to unity, that is $c_s=\rho_0=\mu_0=1$.  Note also that because
 $\rho \simeq  \rho_0$  for our weakly compressible simulations,  the values of $\nu$ quoted in the rest of the paper are the values $\nu=\tilde{ \mu}\over \rho_0$ imposed at the beginning of the simulation.


We employ a triply  periodic box  of dimensionless spatial volume $(2 \pi)^3$  with mesh size of $256^3$ for runs with $\eta=\nu$=0.006, and $288^3$ for runs with $\eta=\nu$=0.001. The forcing function $f$ in the momentum equation is either helical($\nabla\times{\bf f}\propto{\bf f}$) or fully non-helical. ${\bf f}(x,t)=N\,{\bf f_k}(t)\, exp\,[i\,{\bf k}(t)\cdot {\bf x}+i\phi(t)]$, here $N$ is a normalization factor, ${\bf k}(t)$ is the forcing wave number with $|{\bf k}(t)|\sim 5$ ($4.5<|{\bf k}(t)|<5.5$).


\section{Results}
During the HF  phase of the simulations,  the large scale  magnetic helicity  $H_1$
at $k=1$ grows exponentially.
Fig.1(a)-1(c) show  the time evolution of $H_1$ for runs of our two $\eta$ cases.
In the first two panels, the
y-axis of the larger panel
are linear and the y-axis on the  inset  small box is logarithmic.      The solid lines give  the time evolution of  $H_1$ in the helical forcing phase in runs for which there was a preceding  NHF phase.   In simulation time $t$, the NHF was turned off and the HF was turned on at $t=714.153$ for the $\nu=\eta=0.006$ case and at $t=584.724$ for the case
 $\eta=\nu=0.001$.  In Figs.1(a) and 1(b), we have  removed the NHF forcing phases and rescaled the time axis to $t'_{006}=t-714.153$
 for  the case of $\nu=\eta=0.006$   and   $t'_{001}=t-584.724$ for the case of  $\eta=\nu=0.001$.

The dashed lines Figs.1(a) and 1(b)   show  $H_1$ for the cases of HF  without an NHF precursor  phase plotted vs. simulation time. The dashed lines on their own (to the left and right of the solid curves in Fig.1(a) and Fig.1(b) respectively) show  $H_1$ plotted versus simulation time from $t=0$. The dashed lines overlayed on the solid lines represent the same simulation data as the dashed lines plotted on their own, but  time-shifted to aid the visual  comparison to the solid curves.  To  overlay the dashed lines with the solid lines requires shifting them by different amounts for the $\eta=0.006$ and $\eta=0.001$ cases.   These shifts are given respectively by   $t_{006}''=t_{006}'+275$, $t_{001}''=t_{001}'-128$.



The lines in Figs.1(a) and 1(b) are well matched with the time-shifted dashed lines,  highlighting that  the growth rates of $H_1$ in the HF phase are  independent of the presence or absence  a NHF  precursor in both the $\eta=0.001$ and $\eta=0.006$ cases for most of the time evolution.  However, the inset log plots show that the solid and dashed lines for the  $\eta=0.001$  do not evolve so linearly as those 
for the $\eta=0.006$ case at very early times.
This results because the $\eta=0.001$ case incurs a SSD in the NHF  phase
that precedes the HF for the solid curve.  The  larger resistivity for the $\eta=0.006$ case ($Re_M\sim40$) prevents a SSD from occurring in the NHF phase. These differences lead to different value of  the ratio of $E_1/H_1$ at the onset of HF for all of these cases,  where  $E_1$ is the total magnetic energy at $k=1$. We discuss the influence of this ratio in  Section 4.


Fig.1(c) shows  all of the four simulations just described, plotted on the same axes  and then plotted again but time-shifted to  overlap so  that their slopes can be mutually compared.
The deviations from  the curves being exactly aligned is  primarily due to
the different values of $\eta$  rather  than on the presence or absence of a precursor SSD before  LSD growth.

\begin{figure*}

\caption{Before  the $HF$ phase begins at $t=714$, magnetic energy decays
(left small box). But once $HF$  begins, the growth of kinetic energy at $k=5$ dominates  the increase of $E_{kin}$ for $714<t<1200$(right small box). Notation $k=[k_{ini}, k_{max}]$ indicates wave numbers summed in computing the the contributions to $E_{kin}$ or $E_{mag}$.}

{
     \includegraphics[width=14.2cm]{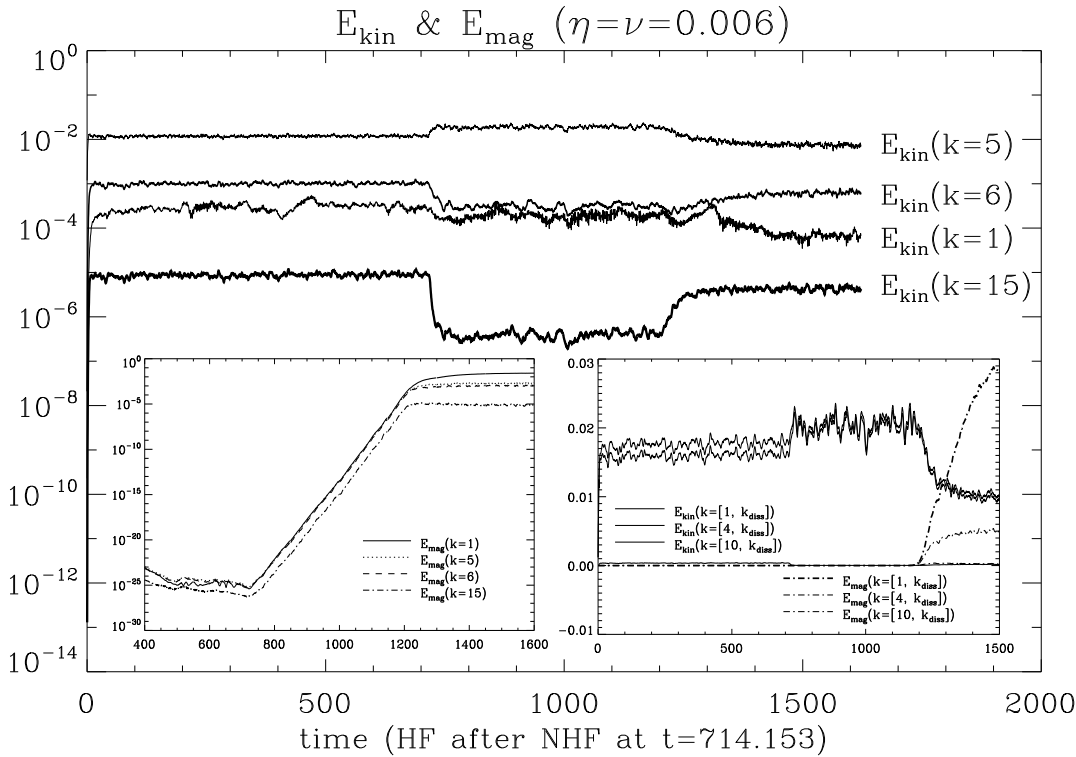}
}

\label{ekem006total5a}
\end{figure*}

For $\eta$(=$\nu$)=0.006  (Fig.2), the simulation is initiated with seed values of  $|\bfv_0|=0$, $|\bfb_0|\sim10^{-2}$(small box in Fig.3(a)). During the NHF phase, amplification of magnetic energy by   field line stretching is unable to overcome  magnetic field decay  so there  is no SSD. At the end of the $NHF$  phase at $t\sim714$, the system has $\bfv\sim0.19$, $\bfb\sim10^{-12}$, and $\ab\sim2.8\times10^{-28}$ (Table 1).  That is, the magnetic quantities have  decayed from their initial values.   Once the HF  forcing begins at $t=714$, the kinetic energy remains
dominated by that at the the forcing scale $k=5$(Fig.2).
For $714<t<1200$,   which corresponds  to the LSD kinematic regime, the growth of $E_{mag}$ is rapid but   observable only in the log plot inset since the starting field value at $t=714$ is so small.
During this phase, the kinetic energy of the forcing scale incurs a large increase(log scale) whereas the kinetic energy at  $k>5$ decreases. For $1200<t<1300$ the kinematic regime ends and the small scale helical magnetic field begins to saturate.  The kinetic energy at wave numbers  $k\ge 5$  recover to  their  pre-$t=714$ values.

For the  $\eta$(=$\nu$)=0.001 case (Fig.3(a), 3(b)) the NHF phase does produce a SSD, so in the NHF phase the magnetic energy grows rather than decays.  The SSD  produces  a power-law tail toward small wave numbers \cite{1968JETP...26.1031K}, \cite{2013MNRAS.429.2469B}  that amplifies some large scale field well above the initial seed values even before the more dominant HF phase of large scale field growth. At the end of the  $NHF$  phase, the system has  $\bfv\sim0.25$, $\bfb\sim0.074$, and $\ab\sim1.15\times10^{-6}$(Table 1).
When the $HF$ subsequently begins from these  starting values, the LSD takes less time to grow  large scale field to saturation than for the $\eta=0.006$ case when there is no SSD precursor.   The growth rate of $H_1$ becomes very close to that of the case without the SSD for most all of the time range as seen in Fig.1(b).

We show that the evolution of the energy spectrum for the $\eta=0.001$ case,  Figs.3(a) shows the evolution of the energy spectrum, indicating how the SSD builds up the very small scale magnetic energy $k>30$  to near equipartition with the kinetic energy. The kinetic energy near the forcing scale dominates the magnetic energy
and overall the SSD produces a magnetic energy which is $\sim 9\%$ of the kinetic energy. This is a small percentage (and smaller than found in \cite{2013MNRAS.429.2469B}) but
 still it does not seem to cause a corresponding  difference in growth rates when the cases with and without the SSD are compared.

Fig.3(b) shows that the kinetic energy remains steadier at the plotted range of wave numbers for the $\eta=0.001$ case   than for the case of $\eta=0.006$ during the corresponding evolution to saturation shown in  Fig.2.

\begin{figure*}
\caption{Energy spectra and energy evolution for the $\eta=0.001$ case (a) $E_{kin}$ and $E_{mag}$ in the small box are the spectral distribution for the initial default seed field of the code. The curve with a spike  is for $E_{kin}$. The large box shows the saturated spectra for the cases of  $NHF$($E_{k0}$, $E_{m0}$) and $HF$ after $NHF$ ($E_{k, sat}$, $E_{m, sat}$) which coincide with those of $HF$ without a precursor $NHF$ phase. The growth of the SSD shows that the NHF has produce an SSD  (b) Before the $HF$ starts at $t=584$, NHF has produces a SSD, amplifying the magnetic energy to $9\%$ of the total kinetic energy.  Note the drop in kinetic energy and further growth of large scale magnetic energy once the HF starts.}
\mbox{%
   \subfigure[]
   {
     \includegraphics[width=13.5cm]{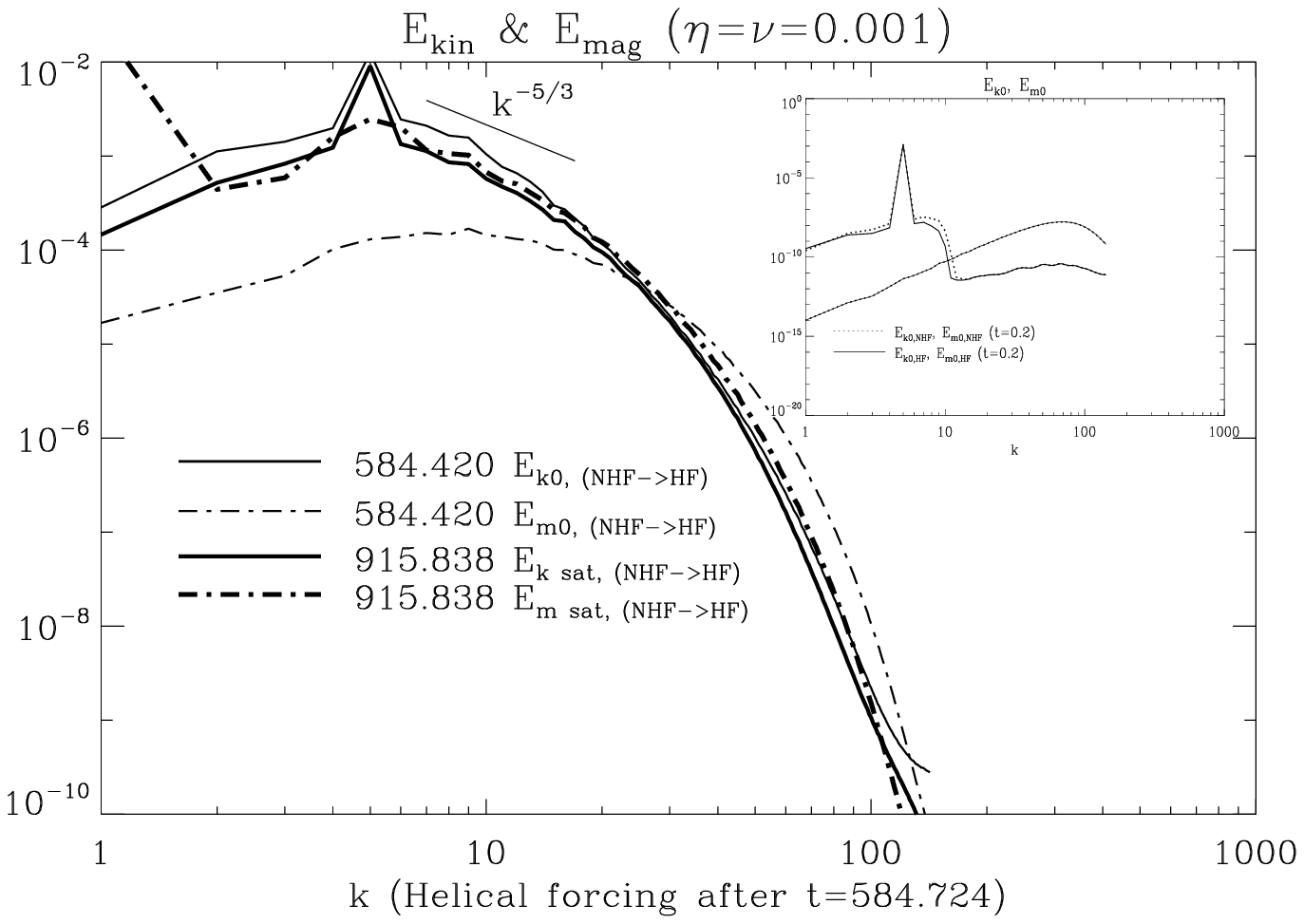}
     \label{ekem001b}
               }
}

\mbox{%
   \subfigure[]
   {
     \includegraphics[width=13.5cm]{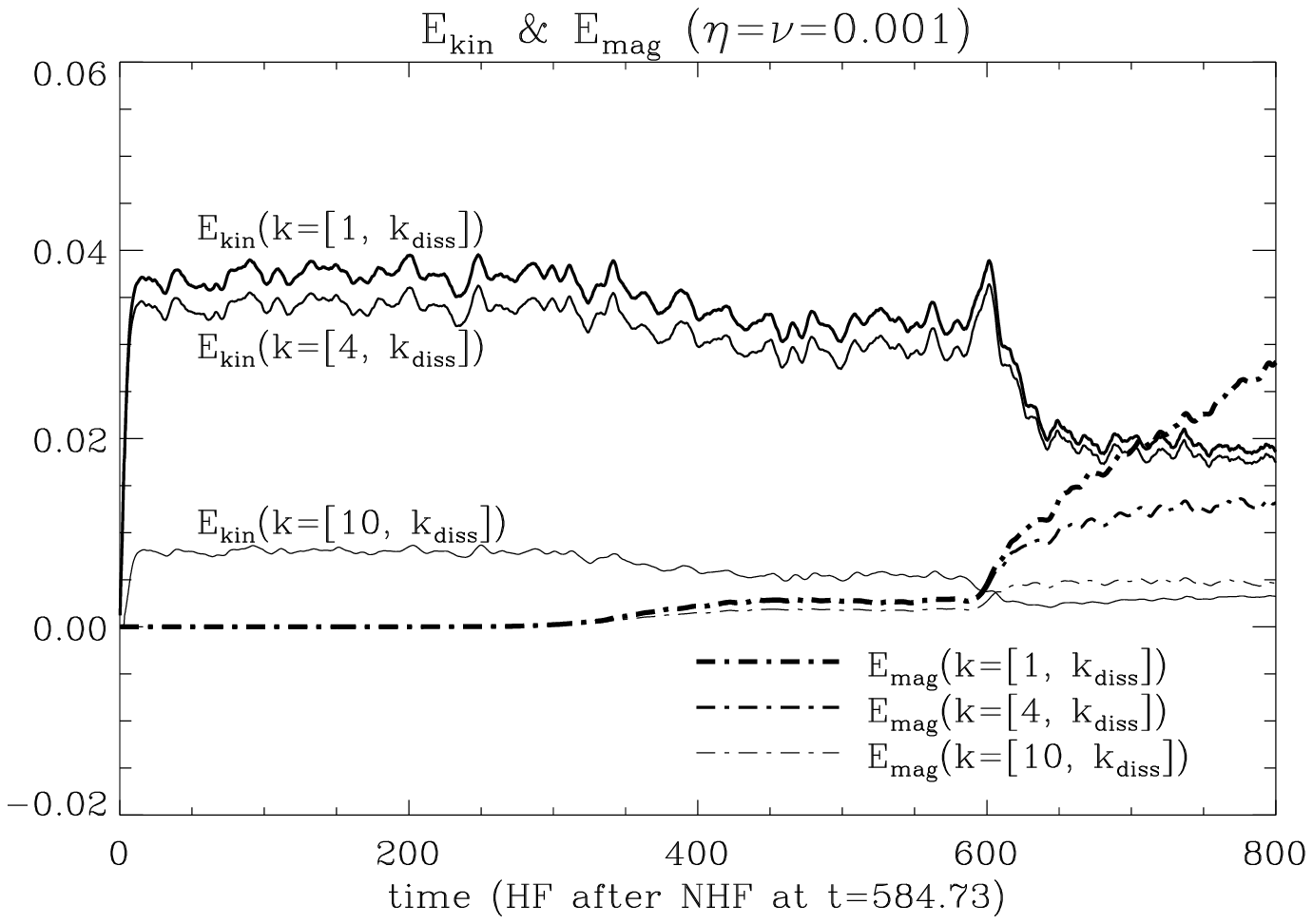}
     \label{ekem001total}
               }
}
\end{figure*}

\begin{table*}

\caption{Initial values in each simulation. Step $\mathrm{I}$: the time range between the start and onset of $HF$. Step $\mathrm{II}$: the time range between onset and saturation of the LSD.}

\begin{tabular}{|c|c|c|c|c|}
\hline
\multirow{2}{*}& $\mathbf{0.006}$($\eta=\nu$) & $\mathbf{0.006}$ & $\mathbf{0.001}$ & $\mathbf{0.001}$\\
& {\small after $NHF$ } & {\small no $NHF$} & {\small after $NHF$} & {\small no $NHF$}\\
\hline

$\mathbf{Step}$&\scriptsize $\bfv_i\sim0.19$, $\bfb_i\sim1\times10^{-12}$& \scriptsize $\bfv_i\sim0$, $\bfb_i\sim0.01$ & \scriptsize $\bfv_i\sim0.25$, $\bfb_i\sim0.074$ & \scriptsize $\bfv_i\sim0$, $\bfb_i\sim1\times10^{-2}$\\
\multirow{1}{*}{$\mathbf{\mathrm{I}}$}& \scriptsize $|H_{Mi}|\sim2.8\times10^{-28}$ & \scriptsize $|H_{Mi}|\sim1\times10^{-14}$  & \scriptsize $|H_{Mi}|\sim1\times10^{-6}$ & \scriptsize $|H_{Mi}|\sim5.4\times10^{-15} $\\
\hline

\multirow{2}{*}{$\mathbf{\mathrm{II}}$}
&\scriptsize $\bfv_i\sim0.19$, $\bfb_i\sim10^{-2}$& \scriptsize $\bfv_i\sim0.18$, $\bfb_i\sim10^{-2}$ & \scriptsize $\bfv_i\sim0.23$, $\bfb_i\sim10^{-2}$ & \scriptsize $\bfv_i\sim0.22$, $\bfb_i\sim10^{-2}$\\
& \scriptsize $|H_{Mi}|\sim10^{-4}$ & \scriptsize $|H_{Mi}|\sim10^{-4}$  & \scriptsize $|H_{Mi}|\sim10^{-4}$ & \scriptsize $|H_{Mi}|\sim10^{-4} $\\
\hline
\end{tabular}
\label{Growth rate table}

\end{table*}



\begin{figure}

\caption{Absolute values of growth rate $\gamma$(=$d H_1/dt$), $E_1/2H_1$, and $\langle\mathbf{v \cdot \omega\rangle}$-$\langle\mathbf{j \cdot b}\rangle$. Thick lines are for $HF$ after $NHF$ and thin lines are for $HF$ without $NHF$. For the comparison, the plots of $HF$ without $NHF$ are shifted and overlaid: (a) $t\rightarrow t+989$ (b) $t\rightarrow t+456$. The lines for $\gamma$ are the mean values of 100 consecutive points. $E_M/2H_1$ and residual helicity are rescaled by multiplication with $2.5\times10^{-4}$. After $\gamma$ arises, there is a time period($t=20\sim 40$) that growth rates of both cases are coincident. (c) After this period, the growth rate of higher resistivity is larger, and drops faster.}

  \subfigure[]
   {
     \includegraphics[width=9cm]{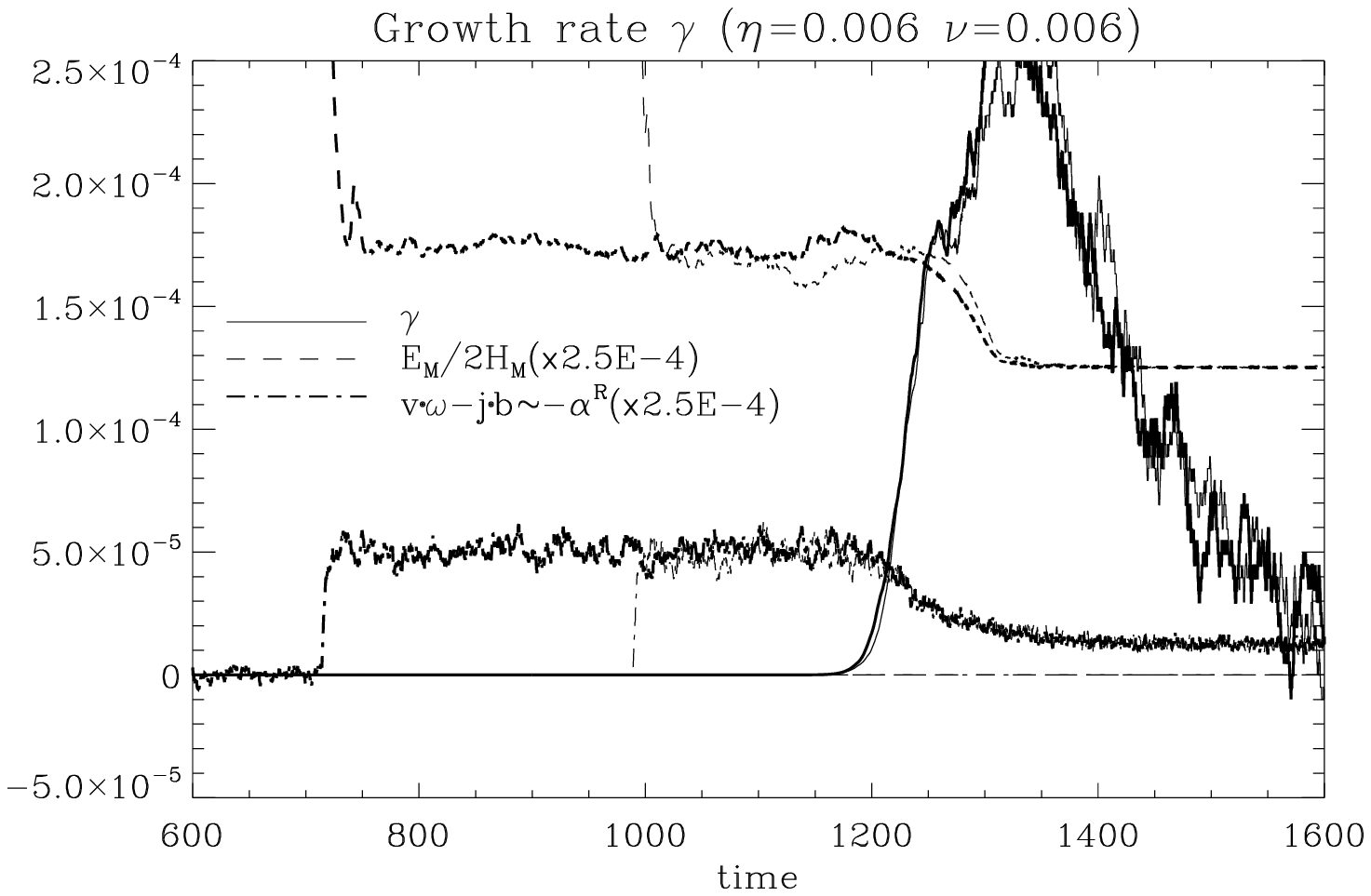}
     \label{growthrate006}
               }
   \subfigure[]
   {
     \includegraphics[width=9cm]{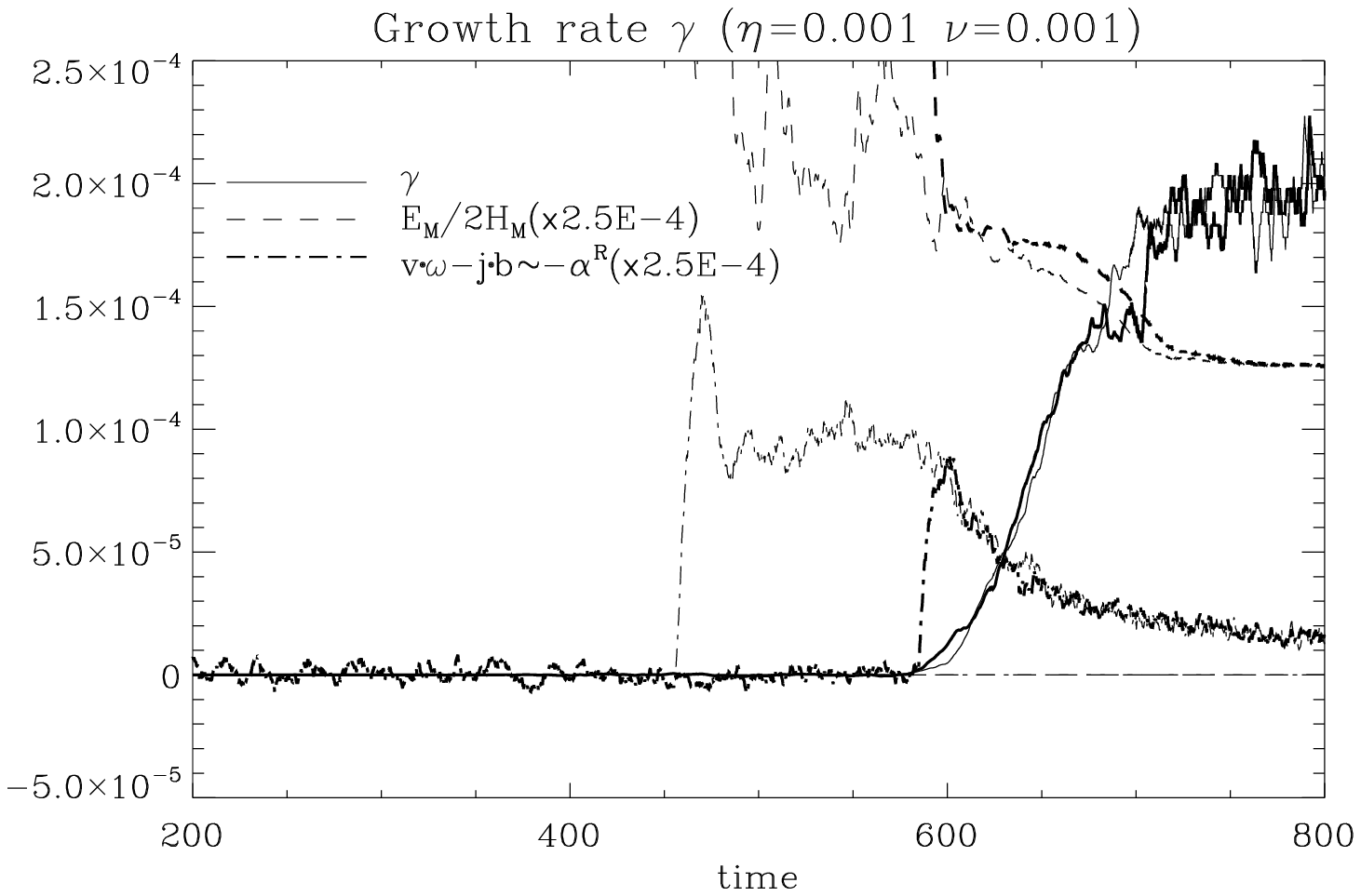}
     \label{growthrate001}
              }
  \subfigure[]
   {
     \includegraphics[width=9cm]{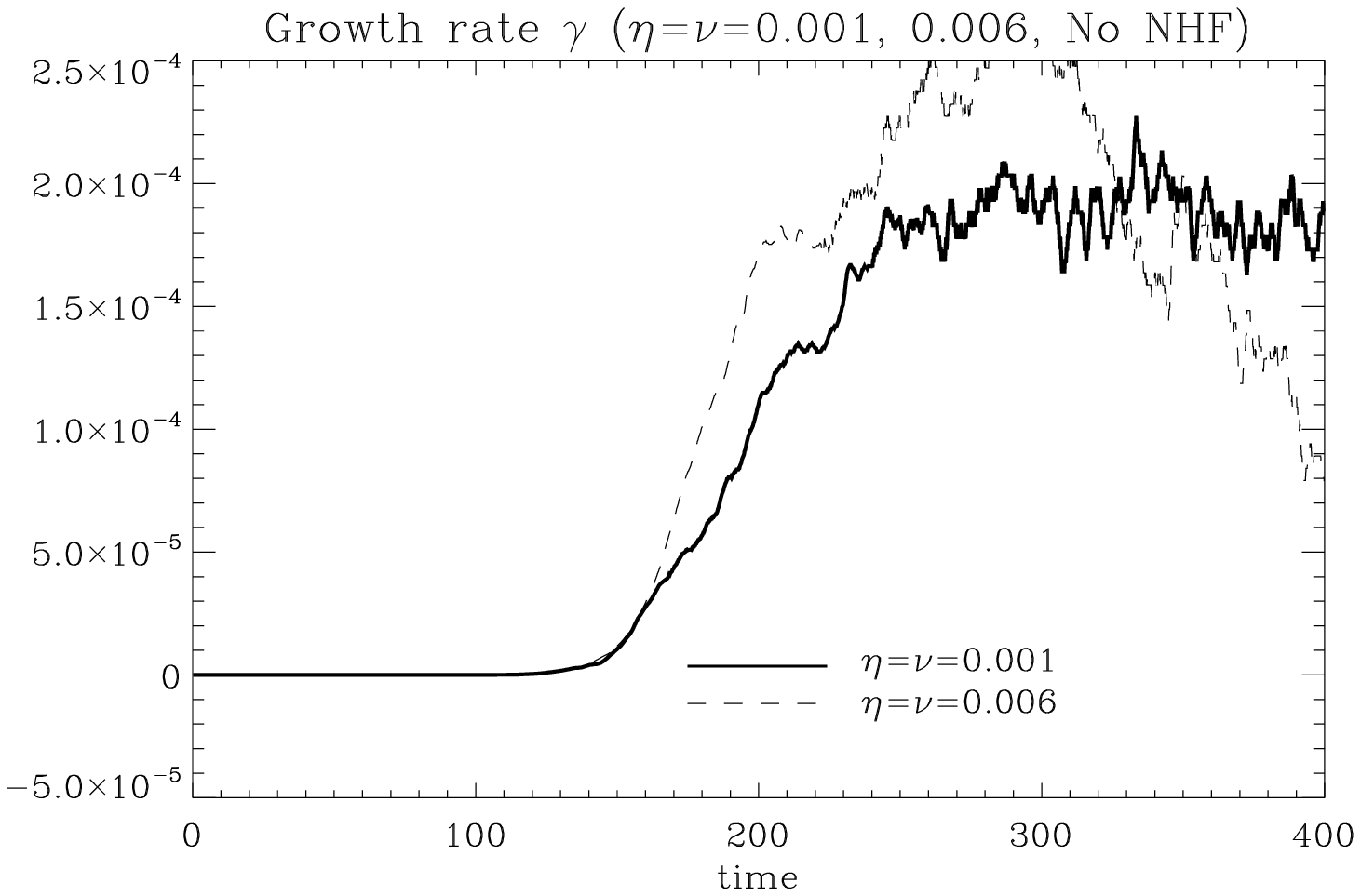}
     \label{growthratecomparison}
               }

\label{growthratefig}
\end{figure}

\section{Assessing the correspondence with two-scale  theory}
Using the  Eddy Damped Quasi Normal Markovian(EDQNM) closure (\cite{1970JFM....41..363O}),  Ref. \cite{1976JFM....77..321P} developed evolution equations for the magnetic energy and helicity spectra
for conditions identical to those we have simulated. Refs. \cite{2002ApJ...572..685F}, \cite{2002PhRvL..89z5007B} showed that under certain limits,  the results from \cite{1976JFM....77..321P},  can be approximated by two-scale approaches. In principle, the simplicity of two-scale approaches gives them practical value so it is important to assess how well they agree with simulations. The ``two-equation''  two scale approach   \cite{2002ApJ...572..685F} involves solving only the time evolution equation for the large scale magnetic helicity $H_1$, and the small scale magnetic helicity $H_2$ assuming $\emfb$ is constant or varies slowly in time compared to other mean quantities. One ``three equation'' two-scale approach involves solving coupled differential equations for $H_1,H_2$ and $\emfb$, the turbulent electromotive force. A different ``three equation'' version of  two-scale theory which solves equations for $H_1,H_2$ and $E_1$ while assuming $\emfb$ is constant has also been studied \cite{2013MNRAS.429.1398B}.  All of these  approaches have the property that the non-helical small scale magnetic energy does not enter the growth rate of the large scale helical field. This specific prediction of these approaches seems consistent with our simulations but we will explain why some features in the simulations seem to require at least a ``four-equation''  version that  incorporates coupled equations for $H_1,H_2, E_1$ and $\emfb$. We now discuss  these points further.

The  equation for  $H_1$ in  two-scale approaches  for which the time evolution of $\emfb$ is assumed constant and the turbulence is assumed to be isotropic is
\beq
\partial_tH_1=
4\alpha E_1 -2(\beta+\eta )k_1^2H_1,
\label{6ab}
\ee
where $k_1$ and
 $k_2$ are the large and small scale wave numbers (the latter assumed to be the forcing scale in a 2-scale model), and
$
\alpha = \tau (k_2^2 H_2- H_{V,2})$
 and 
 $ \beta=\tau E _{V,2}$.
Above, $H_{V,2}$ and $E_{V,2}$ are the kinetic helicity and kinetic energy at $k_2$, and
 $\tau$ has units of time.  Although $\tau$ is taken as a constant $\sim {1\over k_2v_2}$ in the minimalist  two scale theories in more general two-scale theories that require  solving a separate equation for the time derivative of the $\emfb$ \cite{2002PhRvL..89z5007B}
or incorporate memory effects \cite{2009ApJ...706..712H}, \cite{2013MNRAS.428.3569C}, this time
need not be a constant and  emerges from  time integrals.  Different closures can lead to different
interpretations of $\tau$.   We will later see why a  constant $\tau$ is in fact insufficient to explain the simulations.  In addition, we note that in the minimalist ``two-equation'' two-scale theory, $E_1$ is assumed to be maximally helical and equal $ k_1 H_1$. We will see that this is also inconsistent with the simulations. So we keep $E_1$ and $H_1$  separate in the discussion that follows.  We also
note that $E_{V,1}$ the kinetic energy at $k=1$, is likely an important player in amplifying the non-helical magnetic energy on that scale.


The growth rate for the large scale helical field associated with Eq.(\ref{6ab})  is  given by
\begin{eqnarray}
\gamma&=&\frac{1}{H_1}\frac{\partial H_1}{\partial t}=\alpha \frac{E_1}{H_1}-2k_1^2(\beta + \nu_M).
\label{growth rate}
\end{eqnarray}
Initially,  $\alpha$ is dominated by $H_{V,2}$ but eventually the  small scale magnetic helicity $H_2$ grows and
the complete evolution of $H_1$ to saturation  requires also solving the time evolution for $H_2$.

An important prediction of  Eq.(\ref{growth rate})
 is that the non-helical small scale magnetic energy   does not enter any quantity in the equation.
 (This circumstance is somewhat relaxed for the full EDQNM spectral model \cite{1976JFM....77..321P}).
   Our present simulations and analysis  support the result that  the non helical  small scale magnetic energy does not influence the evolution of $H_1$;  the  $H_1$ growth rates for our cases with and without an SSD closely agree.   The small differences we can detect in the $H_1$ growth rates seem to be explicable simply by the different ratios  of $E_1/H_1$ at  the onset of HF in the cases
   with and without NHF because of the different simulation histories up to that time.

Specifically,  Figs.4(a) and 4(b),
 show that for each value of $\eta$, the  growth rate curves  for the separate simulations  of  NHF+ HF and  HF without NHF nearly overlap  on linear plots.  Fig 4(c) shows  that  the value of $\eta$ has more influence than the presence or absence of a precursor NHF phase.
 On log plots we would see a  non-overlap at early times of each of the pairs of growth rate curves   in Figs.4(a) and 4(b)
   corresponding to the difference in slopes of each  pair of plots in the log plot insets  of  Figs.1(a) and 1(b).  (Note that  the plots of   Figs 1(a), 1(b) and
and Figs.4(a), 4(b) all begin at the time corresponding to
the start of the HF phase, despite the different shifted  time axis coordinates).
The different simulation histories of $E_1/H_1$, also shown in Figs.4(a) and 4(b),  can explain these differences:
For  the $\eta=0.001$ case of NHF+HF, the NHF phase produces
a SSD such that at the onset of HF,  larger $E_1$ and $H_1$ than those of the $\eta=0.001$ case without NHF emerges.  For the $\eta=0.006$ case there is no SSD during the NHF phase instead  just an enhanced decay of the seed field. In this case, the case of  NHF+HF begins the HF phase with smaller $E_1$ and $H_1$ than those of the case without NHF.

 The time evolution of ${E_1\over 2H_1}$  in conjunction with the  plots of the residual helicity ${\alpha / \tau}$ shown in Figs.4(a) and 4(b),
  is  important for two other reasons in the assessment of  two-scale theories:  (1) First, since fully helical energy at $k_1=1$ corresponds to ${E_1\over 2H_1}=1/2$ on that plot,  we see that for much of the exponential growth regime, there is a significant fraction of non-helical
  magnetic energy at $k=1$.   For example,
  the inset of    Fig.2 in combination with Fig.4(a)),
  highlights that  without an SSD, ${E_1\over 2H_1}$ exceeds $1/2$ for more than 20 orders of magnitude of growth in $H_1$.    This  cannot be accommodated by the ``two-equation'' version of
  two-scale theory which enforces   $E_1=k_1H_1$   (2) Second, we see that the growth rates in  Figs.4(a)
Figs.4(b) do not go to zero  at the same time that $\alpha\over \tau$  and
${E_1\over 2H_1}$ saturate.  This implies that $\tau$ cannot  be a constant in time,  as in the
specific versions of   two-scale models which do not solve for the time evolution of $\emfb$.


\section{Conclusions, Limitations , and Future Directions}

We have studied the evolution large scale magnetic field growth subject to helical velocity forcing
and studied several aspects of the comparison between these simulations and mean field theories that have not been previously studied. First, we found that  the growth rates of helical large scale fields excited by isotropically helically forced  turbulence were independent of the presence of a precursor   SSD that built up
non-helical magnetic energy  many orders of magnitude larger than in the case without the precursor SSD.  As discussed, this specific prediction is consistent with modern two-scale theories of the helical dynamo in which only the helical magnetic fluctuations should affect the large scale field growth.

However there are two  important caveats to this conclusion.
 First is that for our case of $\eta(=\nu)=0.001$ that produced an SSD
the small scale magnetic energy saturated to  only 9\% of equipartition to the kinetic energy before we started  the HF phase.  Although we  see no evidence there being even a 9\% influence of the small scale field on the $H_1$ growth rate,  it is essential in the future  to study cases in which the SSD phase produces magnetic energy with  a larger fraction of equipartition. This will require larger simulations.  Such computations will be  needed to assess whether the two-scale theory is capturing the correct qualitative result that the non-helical small scale field, even when closer to equipartition,  would not affect the LSD growth rate.
In this respect, we regard our present work as identifying the importance of answering this question
and providing some preliminary results toward that end.  We note that even in $512^3$ simulations \cite{2013MNRAS.429.2469B}, the SSD saturates at only ~23\% of equipartition of the kinetic energy.
It is not yet clear whether the saturation percentage  evolves  to a limiting value with increased numerical resolution.

The second caveat is  that in the present simulations we  turned off the NHF once the fully HF started.
In the future,   studying the correspondence between LSD theory and simulation  for fractionally helical forcing  \cite{2002ApJ...566L..41M} is desirable because fractionally helical forcing can be thought of as  essentially  contemporaneous  forcing of both non-helical SSD and helical LSD. The two-scale theory predicts that the non-helical small scale magnetic energy should not affect the large scale growth rate in this case either.

 While our results assessing the influence of an SSD (albeit a weak SSD) did not show any inconsistency with the expectations of two scale mean field theory so far, we did find several features of the large scale field
 growth that cannot be accommodated by the minimalist versions of two scale theories that solve only
 coupled equations for $H_1$ and $H_2$.  Namely, even with fully helical forcing of the velocities at $k=5$, the large scale field at $k=1$ incurs significant non-helical field growth, so a separate equation for $E_1$ is necessary.   Moreover, to source the non-helical magnetic energy, inclusion of
 kinetic energy $E_{k,1}$ on the large scale is seemingly necessary, otherwise the non-helical magnetic energy  at $k=1$ can only decay in the two-scale theory.
  Second, the saturation of the large scale helical field growth is delayed from
 the prediction of a constant $\tau$ in Eq.(\ref{growth rate}). This requires at minimum
 a fourth equation for the time evolution of  $\emfb$ to be coupled into the theory.
 The extent to which a ``four equation'' version of two-scale theory can accommodate these features, or whether the full power of a more comprehensive spectral model \cite{1976JFM....77..321P}
   is required, remain  topics for further work.

\vfill

\bibliography{bibdatabase}
\bibliographystyle{plain}

\end{document}